\documentclass[preprint, amsmath,amssymb,nofootinbib]{revtex4}

\usepackage{graphicx, bm, color}
\usepackage[]{caption}
\usepackage{subfig}

\usepackage{dcolumn}% Align table columns on decimal point
\def\be{\begin{equation}}
\def\ee{\end{equation}}
\def\bea{\begin{eqnarray}}
\def\eea{\end{eqnarray}}
\def\dprime{\prime\prime}

\def\caln{\mathcal{N}}
\def\cals{\mathcal{S}}
\def\mpl{M_{P}}

\def\super3{{}^{(3)}}

\def\tri{\triangle}

\def\caln{\mathcal{N}}
\def\cala{\mathcal{A}}
\def\calb{\mathcal{B}}

\def\calp{\mathcal{P}}

\begin{document}

\title{Relic gravitational wave spectrum, the
trans-Planckian physics and Ho\v{r}ava-Lifshitz gravity}

%\author{Ron-Gen Cai} \email[email: ]{cairg@itp.ac.cn}
%\author{Bin Hu} \email[email: ]{hubin@itp.ac.cn}
\author{Seoktae Koh} \email[email: ]{skoh@itp.ac.cn}

\address{Institute of Theoretical Physics, Chinese Academy of Sciences,
P.O. Box 2735, Beijing, 100080, China}


\begin{abstract}
We calculate the spectrum of the relic gravitational wave 
 due to the trans-Planckian
effect in which the standard linear dispersion relations may be modified. 
Of the modified dispersion relations suggested in literatures
which have investigated the trans-Planckian effect,
we especially use the Corley-Jacobson dispersion relations. The Corley-Jacobson
type modified dispersion relations can be obtained  from
Ho\v{r}ava-Lifshitz gravity which is non-relativistic and UV complete.
Although it is not clear how the transitions from Ho\v{r}ava-Lifshitz
gravity in the UV regime to Einstein gravity in the IR limit occur,
we assume Ho\v{r}ava-Lifshitz gravity regime is followed by the
inflationary phase in Einstein gravity.
\end{abstract}

\maketitle

%%%%%%%%%%%%%%%%%%%%%%%%%%%%%%%%%%%%%%%%%%%%%%%%%%%%%%%%%%%%%%%%%%%%%%%%%
%%%%%%=================INTRODUCTION=======================================
\section{Introduction}
%%%%%%%%%%%%%%%%%%%%%%%%%%%%%%%%%%%%%%%%%%%%%%%%%%%%%%%%%%%%%%%%%%%%%%%%%%

In spite of the success of the inflationary scenario, inflation 
is still encountered several unsolved problems. Among them,
the trans-Planckian problem \cite{Martin:2000xs}
 is related to the fundamental problem
of the  high energy physics {\it i.e} quantum theory of 
gravity. Unfortunately there is 
no successful quantum gravity theory for 
 handling the spacetime around the Planck scale, so one way out of such 
difficulties is to use the effective modified
dispersion relations to mimic the effect of the trans-Planckian 
physics. Of the several modified dispersion relations suggested in
many literatures, we will use the Corley-Jacobson dispersion 
relation \cite{Martin:2000xs}\cite{Corley:1996ar} in this work.

The Corley-Jacobson type dispersion relations could be derived naturally 
from the recently proposed quantum gravity model called
 Ho\v{r}ava-Lifshitz(HL) gravity \cite{Horava:2009uw}\cite{Horava:2008ih}. 
HL gravity is 
a non-relativistic and renormalizable theory in the UV region.
In addition, because of the different scaling behaviors of the time and space
($t\rightarrow l^3 t,~ {\bf x} \rightarrow l {\bf x}$), 
while the kinetic term is still quadratic order in time derivatives
of the metric in the action,
the potential terms may have  up to 6th order spatial derivatives 
\cite{Calcagni:2009ar}\cite{Kiritsis:2009sh}.
HL gravity may also provide the seeds for the large scale structure 
without inflation \cite{Kiritsis:2009sh}\cite{Mukohyama:2009gg}. 
On the other hand,
it is still controversial if HL gravity is a complete 
theory\cite{Li:2009bg} which would be dependent upon the
projectability and the detail balance condition \cite{Horava:2009uw}.
No explicit inconsistencies has been pointed out against the version
without the detailed balance condition with the projectability condition
\cite{Mukohyama:2009mz}.
And it is also not clear
how the transition from HL gravity  in the UV regime  
to Einstein gravity in the IR limit takes place.

In this paper, we calculate the spectrum of the relic gravitational wave
in order to investigate the effect of the trans-Planckian physics. 
Especially, we use the Corley-Jacobson dispersion relations taking
HL gravity into account. 
For this purpose, we assume the HL gravity phase is followed by the inflationary
phase in Einstein gravity.
Though the transition time from
the trans-Planckian regime to the usual standard inflationary period
is usually determined by the cut-off momentum, 
we will leave it as a free parameter 
assuming it keep the information about the transition mechanism.
The resulting spectrum could be constrained from the
experimental bounds such as  the 
ground based gravitational wave detector (LIGO, VIRGO) which covers
the frequency range $ 10\text{Hz} < f < 10^3 \text{Hz}$ and future planned
space based one (LISA) which is expected to cover $10^{-3}\text{Hz}<
f < 1 \text{Hz}$ or the cosmological bounds (nucleosynthesis, CMB)
in the low frequency range \cite{Maggiore:1999vm}. 

The paper is organized as follows. In Sect. \ref{sect:gwtrans}, 
we calculate the power spectrum of the tensor perturbations with the
Corley-Jacobson dispersion relations. In Sect. \ref{sect:gwhl}, we 
obtain the equation for the tensor perturbations in HL gravity and
calculate the power spectrum. In Sect. \ref{sect:gwspect}, we investigate
the effect of the trans-Planckian  physics and HL gravity 
on the spectrum of the relic gravitational wave. And finally, we summarize
in Sect. {\ref{sect:summary}.

%%%%%%%%%%%%%%%%%%%%%%%%%%%%%%%%%%%%%%%%%%%%%%%%%%%%%%%%%%%%%%%%%%%%%%%%%%
%%%%%%%================GWs=======================================
\section{Gravitational wave spectrum in the trans-Planckian physics}
\label{sect:gwtrans}
%%%%%%%%%%%%%%%%%%%%%%%%%%%%%%%%%%%%%%%%%%%%%%%%%%%%%%%%%%%%%%%%%%%%%%%

We briefly review the effect of  the trans-Planckian physics 
on the power spectrum in this section.
With the metric
\bea
ds^2 = a^2(\eta) [ -d\eta^2 + (\delta_{ij} + h_{ij})dx^i dx^j ],
\label{eq:metric}
\eea
where $h_{ij}$ is a transverse, traceless perturbation which can be
expanded as
\bea
h_{ij} (x) = \frac{1}{a}\int \frac{d^3 k}{(2\pi)^{3/2}}
\sum_{ij} \epsilon_{ij} u_k (\eta) e^{i {\bf k}\cdot {\bf x}},
\label{eq:modexp}
\eea
and $\epsilon_{ij}$ is a polarization tensor, 
the tensor perturbations are governed by the equations
\bea
u_k^{\dprime} + \biggl(k^2 - \frac{a^{\dprime}}{a} \biggr) u_k = 0,
\label{eq:stdeq}
\eea
where a prime denotes a derivative with respect to the conformal time 
and $u_k$ satisfy the normalization condition
\bea
u_k u_k^{\ast \prime} - u_k^{\ast} u_k^{\prime} = i.
\label{eq:normcond}
\eea
This equation is believed to hold only at the low energy scale, 
$k_{phys} \ll M_{pl}$, where $k_{phys} = k/a$.
%it is not likely to hold still in the high energy scale.
%But 
As we go back in time around the Planck scale,
quantum effects would be important and it seems to be necessary introducing
the quantum gravity theory 
in order to describe correctly the Planck scale phenomena.
Unfortunately, however,  no successful quantum gravity theory exist at present.
Though the string theory is believed to be a strong candidate of the 
quantum gravity theory, it is not complete yet. 
Thus, one way to overcome these difficulties and to mimic 
the quantum gravity effects around the Planck scale is
to consider  the modified dispersion relations.

Taking into account of the modified dispersion relations, $\omega(k/a)$, 
the equation for $u_k$ could have the following form
 %with the modified dispersion relation, $\omega(k)$
\bea
u^{\dprime}_k + \biggl( \omega^2(k/a) - \frac{a^{\dprime}}{a} \biggr) u_k = 0.
\label{eq:modeq}
\eea
%is assumed to be true to describe around the Planck scale.

In this paper, we take the Corley-Jacobson dispersion 
relations\cite{Corley:1996ar}\cite{Martin:2000xs}
as an example of the modified dispersion relations
\bea
\omega^2(k/a) = k^2 \biggl( 1 + b_m \biggl(\frac{k}{aM}\biggr)^{2m} \biggr),
\eea
where $M$ is a cutoff scale and 
$b_m$ is an  arbitrary coefficient.  Here $k$ is a comoving wave number.
Depending on the sign of $b_m$ which  can be positive 
or negative \cite{Martin:2000xs}, we may reach to the different results.
  If $b_m < 0$, $\omega^2$ becomes a complex function for $k > |b_m|^{-1/2m}aM$
and  the WKB approximation may not be valid \cite{Starobinsky:2001kn}
in this region.
For the purpose of this work, bearing in mind HL gravity, it is enough
to limit to the positive $b_m$. From now on we only consider $b_m > 0$.
(For $b_m < 0$, see Ref. \cite{Martin:2000xs}).

If we assume exponential inflation
\bea
a(\eta) = -\frac{1}{H\eta},
\label{eq:scaleinf}
\eea
the equation (\ref{eq:modeq}) yields
\bea
u_k^{\dprime} + \biggl( k^2 + \sigma^2(k) k^{2m+2}\eta^{2m} 
- \frac{2}{\eta^2} \biggr)
u_k = 0,
\label{eq:transmodeq1}
\eea
where 
\bea
\sigma^2 (k) =  b_m \biggl(\frac{H}{M} \biggr)^{2m}.
\eea

For $k> aM$, since $\omega^2(k/a) \approx \sigma^2 k^{2m+2}\eta^{2m}$,
the equation (\ref{eq:transmodeq1}) becomes
\bea
u_k^{\dprime} + \biggl( \sigma^2 k^{2m+2} \eta^{2m} 
- \frac{2}{\eta^2} \biggr) u_k =0,
\label{eq:transmodeq2}
\eea
and its solutions are given by 
\bea
u_k = c_1 \sqrt{\frac{\pi}{4(1+m)}}|\eta|^{1/2} H_{\nu}^{(1)} (z) 
+ c_2\sqrt{\frac{\pi}{4(1+m)}} |\eta|^{1/2} H_{\nu}^{(2)}(z),
\eea
where $H_{\nu}^{(1)}$ and $H_{\nu}^{(2)}$ are the Hankel function of
the first and second kind with the order $\nu$, respectively, and
\bea
z= \frac{\sigma}{1+m}(k\eta)^{1+m}, \quad \nu = \frac{3}{2(1+m)}.
\eea
For $b_m < 0$, the solutions are given by the modified Bessel functions.

If we choose the positive frequency modes as initial conditions 
when $k> aM$, we can choose $c_1 = 0, ~c_2 = 1$ taking 
 the normalization  condition (\ref{eq:normcond}) into account.
This choice of the initial vacuum is equivalent to  the
minimizing the  energy density\cite{Martin:2000xs} up to the phase factor
\bea
u_k(\eta_i) = \frac{1}{\sqrt{2\omega}}, \quad
u_k^{\prime}(\eta_i) = -i\sqrt{\frac{\omega}{2}},
\eea
if the adiabatic approximation ($|\omega^{\prime}| \ll \omega^2$) is hold.
Then the mode solutions of (\ref{eq:transmodeq2})
 for $k> aM $ can be written as
\bea
u_k &=& \sqrt{\frac{\pi}{4(1+m)}}|\eta|^{1/2} H_{\nu}^{(2)}(z) .
\eea

For $k < aM$, the dispersion relations
recover the standard linear relations, $\omega^2 \simeq k^2$,
as in (\ref{eq:stdeq}) and then the mode
solutions are given by
\bea
u_k (\eta) 
%&=& \sqrt{\frac{\pi}{4}}|\eta|^{1/2} [ \cala_k H_{3/2}^{(1)} (k|\eta|)
%+ \calb_k H_{3/2}^{(2)}(k|\eta|) ] \\
&=& \frac{\cala_k}{\sqrt{2k}}\biggl( 1- \frac{i}{k\eta}\biggr) 
e^{-i k\eta} + \frac{\calb_k}{\sqrt{2k}}\biggl( 1 + \frac{i}{k\eta}\biggr) 
e^{i k\eta},
\label{eq:infsol}
\eea
where the coefficients satisfy the normalization condition
\bea
|\cala_k|^2 - |\calb_k|^2 = 1.
\eea

%After the end of inflation, the mode function becomes
%\bea
%u_k(\eta) = \frac{\alpha_k}{\sqrt{2k}} e^{- ik\eta} 
%+ \frac{\beta_k}{\sqrt{2k}} e^{ ik\eta},
%\eea
%and
%\bea
%|\alpha_k|^2 - |\beta_k|^2 = 1.
%\eea

The coefficients are determined through the matching conditions at 
$\eta = \eta_1$ where 
$\eta_1$ is the time when the modified dispersion relations take 
the standard linear form,
\bea
k|\eta_1| = \frac{1}{b_m^{1/2m}} \frac{M}{H} \gg 1.
\label{eq:eta1}
\eea

Through the matching conditions at $\eta = \eta_1$,
 we obtain the coefficients
\bea
\cala_k &=& \sqrt{\frac{\pi}{8(1+m)}} (k|\eta_1|)^{1/2}
\biggl( 1+\frac{3i}{2k|\eta_1|} -\frac{3}{2k^2 |\eta_1|^2} \biggr)e^{ik|\eta_1|}
H_{\nu}^{(2)} (z_1) 
\nonumber \\
& & + \sqrt{\frac{\pi(1+m)}{8}}\frac{iz_1}{(k |\eta_1|)^{1/2}}
\biggl( 1+ \frac{i}{k|\eta_1|} \biggr) e^{ik|\eta_1|} \frac{d}{dz}H_{\nu}^{(2)}(z_1), 
\\
\calb_k &=& \sqrt{\frac{\pi}{8(1+m)}}(k |\eta_1|)^{1/2}
\biggl( 1 -\frac{3i}{2k|\eta_1|} - \frac{3}{2k^2 |\eta_1|^2} \biggr)
e^{-ik |\eta_1|} H_{\nu}^{(2)}(z_1)
\nonumber \\ 
& & -\sqrt{\frac{\pi(1+m)}{8}} \frac{iz_1}{(k |\eta_1|)^{1/2}}
\biggl(1- \frac{i}{k|\eta_1|} \biggr) e^{-ik|\eta_1|} \frac{d}{dz} H_{\nu}^{(2)}(z_1),
\eea
where
\bea
z_1 = \frac{\sigma}{(1+m)} (k|\eta_1|)^{1+m}.
\label{eq:z1}
\eea

With these coefficients,
 we can compute the power spectrum of the gravitational wave
at $\eta \rightarrow 0$ or $k \eta \ll 1$
\bea
\calp_h &=& \frac{1}{a^2} \calp_u =  \frac{k^3}{2\pi^2}\frac{|u_k|^2}{a^2}
\nonumber \\
&=& \biggl(\frac{H}{2\pi} \biggr)^2 \biggl( 1 + 2|\calb_k|^2 
-Re(\cala_k \calb_k^{\ast}) \biggr),
\label{eq:power}
\eea
where we have used $a = -\frac{1}{H\eta}$ at the second line.

%\bea
%|\calb_k|^2 &=& \frac{1}{4(1+m)}\frac{k\eta_1}{z_1}
%\biggl[1- \frac{(2+m)(2-m)}{4 k^2 \eta_1^2} + \frac{(2-m)^2}{4k^4 \eta_1^4}
%-\frac{2(1+m)z_1}{k\eta_1} 
%\nonumber \\
%& & + \frac{(1+m)^2 z_1^2}{k^2\eta_1^2}
%+\frac{(1+m)^2 z_1^2}{k^4 \eta_1^4} \biggr],  \\
%\cala_k \calb_k^{\ast} &=& \frac{1}{4(1+m)}\frac{k\eta_1}{z_1}
%e^{2ik\eta_1} \biggl[ 1 - \frac{(m-6)(m-2)}{4k^2 \eta_1^2}
%+\frac{(2-m)^2}{4k^4 \eta_1^4} - \frac{(1+m)^2 z_1^2}{k^2 \eta_1^2}
%\nonumber \\
%& & +\frac{(1+m)^2 z_1^2}{k^4 \eta_1^4} 
%+\frac{2i}{k\eta_1} \biggl\{ \frac{2-m}{2}\biggl(1-\frac{2-m}{2k^2\eta_1^2}
%\biggr) - \frac{(1+m)^2z_1^2}{k^2 \eta_1^2} \biggr\} \biggr].
%\label{eq:coeffir}
%\eea

For $k|\eta_1| \gg 1$, using (\ref{eq:eta1}) and (\ref{eq:z1}),
we finally obtain
\bea
\calp_h(k) &\approx&  \biggl(\frac{H}{2\pi} \biggr) 
\biggl( 2 - \frac{b_m^{1/2m}}{4} \frac{H}{M} \sin (2k|\eta_1|) \biggr),
\label{eq:transpect}
\eea
where we have used the asymptotic form of the Hankel function
\bea
H_{\nu}^{(2)}(z) \simeq \sqrt{\frac{2}{\pi z}} e^{-i(z-(\nu+\frac{1}{2})
\frac{\pi}{2})}, \quad \text{if}~~ z \gg 1,
\label{eq:asymphankel}
\eea
and
\bea
\frac{d}{dz}H_{\nu} = -H_{\nu+1} + \frac{\nu}{z} H_{\nu}. \nonumber
\eea

The power spectrum (\ref{eq:transpect}) is the well-known 
result of the trans-Planckian effect.
The leading correction term due to the modified dispersion relations
leads to $\frac{H}{M}$. The detectability of this effect is an
important concern at the future experiments.

%When we derive this result, we use the time $\eta_1$ which is
%dependent upon $k$. 

%Instead of $k$-dependent $\eta_1$, if we choose the constant
%$\eta_1$ which means all modes are matched at the same time.
%Then the power spectrum is much different from the previous result.

%%%%%%%%%%%%%%%%%%%%%%%%%%%%%%%%%%%%%%%%%%%%%%%%%%%%%%%%%%%%%%%%%%%%%
%%%%%%%%%%%%%%%%%%% GWs in HL gravity %%%%%%%%%%%%%%%%%%%%%%%%%%%%%%%
\section{Gravitational wave in Ho\v{r}ava-Lifshitz gravity}
\label{sect:gwhl}
%%%%%%%%%%%%%%%%%%%%%%%%%%%%%%%%%%%%%%%%%%%%%%%%%%%%%%%%%%%%%%%%%%%%

The quadratic action of
Ho\v{r}ava-Lifshitz gravity with the metric (\ref{eq:metric}) is
\bea
\delta_2 \cals_T &=& \int dt d^3x a^2 \biggl[ \frac{2}{\kappa^2} h_{ij}^{\prime}
h^{ij \prime}  + \alpha_1 h^{ij}\tri h_{ij} + \frac{\alpha_2}{a^2}\tri h^{ij} \tri h_{ij}
\nonumber \\
& & +\alpha_3 \frac{\epsilon^{ijk}}{a^3}\tri h_{il} \nabla_j\tri h^l_k
-\frac{\alpha_4}{a^4} \tri h^{ij} \tri^2 h_{ij} \biggl],
\label{eq:quadaction}
\eea
where $\nabla_i$ is a covariant derivative with respect to $\delta_{ij}$
and  $\tri = \delta^{ij} \nabla_i \nabla_j$.
The coefficients in the action corresponding to the original 
theory \cite{Horava:2009uw} are
\bea
\alpha_1 &=& \frac{\kappa^2 \mu^2}{8(1-3\lambda)} \Lambda, 
\quad \alpha_2 = -\frac{\kappa^2 \mu^2}{8},
\quad  \alpha_3 = \frac{\kappa^2 \mu}{2w^2}, 
\quad \alpha_4 = -\frac{\kappa^2}{2w^4}, 
\eea
where $\kappa, \lambda, w$ and $\mu$ are coupling constants with
scaling dimensions $[\kappa] = [w] = [\lambda] = 0, ~[\mu] = 1$
and $\Lambda$ is the 3-dimensional cosmological constant with 
scaling dimension $[\Lambda] = 2$.
 Einstein gravity
is recovered if $\lambda$ is approaching unity in the IR limit.

The background Friedmann equation in HL gravity \cite{Horava:2009uw}
is given by \cite{Calcagni:2009ar}
\bea
H^2 \equiv \frac{\dot{a}^2}{a^2}
= \frac{\kappa^2}{6(3\lambda-1)}a^2 (\rho - \sigma),
\eea
where $\rho$ is the background energy density of the matter and
\bea
\sigma = - \frac{\kappa^2 \mu^2}{8(1-3\lambda)} 3\Lambda^2.
\eea
If $\rho=0$, there is no cosmological solution 
\cite{Calcagni:2009ar}\cite{Minamitsuji:2009ii}. In order to get
de Sitter solutions, we need to introduce
the matter where we assume the vacuum energy density of the matter 
is dominant. In what follows, we assume $\rho_0 \equiv \Lambda_m > \sigma$,
%for the existence of the de Sitter solution. Then, during UV limit,
and then the scale factor is given by
\bea
a = - \frac{1}{H\eta}.
\eea

Note that by the analytic continuation of the parameters $\mu \rightarrow
i\mu$ and $w^2 \rightarrow i w^2$ \cite{Lu:2009em}, the de Sitter 
solutions can be obtained without introducing the 
matter \cite{Minamitsuji:2009ii}.
The sign of the coefficients in the transformed action is also changed.
In order to couple matter with gravity consistently, the relativistic
formulation of HL gravity are also investigated \cite{Germani:2009yt},
in which, although  being fully-relativistic, it results to be locally
anisotropic in the time-like and space-like directions defined by a family
of  irrotational observers.

In the action (\ref{eq:quadaction}), 
because of $\alpha_3$ term, the tensor perturbations are not
invariant under the parity transformation (${\bf x} \rightarrow 
-{\bf x}$) \cite{Takahashi:2009wc}. The chirality of the gravitational
wave seems to give rise to the
 important effect for discriminating HL gravity
from the other quantum gravity theories at the 
future experiment  \cite{Takahashi:2009wc}\cite{Cai:2007xr}. But since
the $\alpha_3$ term is subdominant compared to the $\alpha_4$ term
in the UV limit,  we will drop such term in the present work. 

Varying the action (\ref{eq:quadaction})  leads to the equation
for the tensor perturbations in HL gravity
\bea
u_k^{\dprime}  + \biggl[ \alpha_1 \kappa^2 k^2 \biggl( 1
+ \frac{|\alpha_2|}{\alpha_1 a^2}k^2 + \frac{|\alpha_4|}{\alpha_1 a^4}k^4 \biggr)
 - \frac{2}{\eta} \biggr] u_k  = 0,
\eea
where we have used the mode expansion (\ref{eq:modexp}).
In the UV limit, the dispersion relation of this equation 
is exactly equivalent to the 
Corley-Jacobson  dispersion relation if $m=2$.
If only the dominant term is kept with putting
\bea 
M^4 = \frac{1}{\kappa^2 |\alpha_4|}, \quad b_m = 1,
\eea
then the positive mode solution is
\bea
u_k = \sqrt{\frac{\pi}{12}}|\eta|^{1/2} H_{1/2}^{(2)}(z),
\quad z = \biggl(\frac{H}{\sqrt{3}M}\biggr)^2 (k\eta)^3.
\eea
Since the Hankel function with the order $\nu= 1/2$ can be 
expressed as
\bea
H_{1/2}^{(2)}(z) = i\sqrt{\frac{2}{\pi z}} e^{-iz},
\eea
the power spectrum of the tensor perturbations in HL gravity
\bea
\calp_u(k) \propto k^3 |u_k|^2 \propto k^0
\eea
can be scale invariant in the UV region. If HL gravity can provide 
the seeds for the large scale structure without 
the inflationary phase as was suggested in 
\cite{Kiritsis:2009sh}\cite{Mukohyama:2009gg}, it would also be interesting
to find some observational signature due to the tensor perturbations
in HL gravity  without introducing the inflationary phase \cite{koh09}.
But in this work, we assume HL gravity in the UV region is followed by
the inflationary phase in Einstein gravity.

We, then,  move to the IR limit.
Since it is expected Einstein gravity would be recovered 
in the IR limit with $\lambda = 1$, 
the equation for the tensor perturbations takes the standard form
as like in(\ref{eq:stdeq}).
The solutions, then,  are
\bea
u_k = \frac{\cala_k}{\sqrt{2k}}\biggl( 1- \frac{i}{k\eta}\biggr) 
e^{-i k\eta} + \frac{\calb_k}{\sqrt{2k}}\biggl( 1 + \frac{i}{k\eta}\biggr) 
e^{i k\eta}.
\eea
As in Sect. \ref{sect:gwtrans}, we use the matching conditions 
at $\eta = \eta_1$ in order to determine
the coefficients $\cala_k$ and $\calb_k$. Although it is unclear
how the transition from HL gravity in the UV region 
to Einstein gravity in the IR limit occurs, we assume
that the transition occurs instantaneously at $\eta = \eta_1$.
%However, I would like to stress the importance of 
%the transition mechanism which 
%would be give an effect on the observable power spectrum.

We, then, obtain the coefficients via  the matching conditions
\bea
\cala_k &=& \sqrt{\frac{\pi}{24}} (k|\eta_1|)^{1/2}
\biggl( 1+\frac{3i}{2k|\eta_1|} -\frac{3}{2k^2 |\eta_1|^2} \biggr)e^{ik|\eta_1|}
H_{1/2}^{(2)} (z_1) 
\nonumber \\
& & + \sqrt{\frac{3 \pi}{8}}\frac{iz_1}{(k |\eta_1|)^{1/2}}
\biggl( 1+ \frac{i}{k|\eta_1|} \biggr) e^{ik|\eta_1|} \frac{d}{dz}H_{1/2}^{(2)}(z_1), 
\\
\calb_k &=& \sqrt{\frac{\pi}{24}}(k |\eta_1|)^{1/2}
\biggl( 1 -\frac{3i}{2k|\eta_1|} - \frac{3}{2k^2 |\eta_1|^2} \biggr)
e^{-ik |\eta_1|} H_{1/2}^{(2)}(z_1)
\nonumber \\ 
& & -\sqrt{\frac{3\pi}{8}} \frac{iz_1}{(k |\eta_1|)^{1/2}}
\biggl(1- \frac{i}{k|\eta_1|} \biggr) e^{-ik|\eta_1|} \frac{d}{dz} H_{1/2}^{(2)}(z_1),
\eea
where
\bea
z_1 = \biggl(\frac{H}{\sqrt{3}M}\biggr)^2 (k|\eta_1|)^3.
\eea

Since, for $k \eta \ll 1$,
\bea
|\calb_k|^2 &=& \frac{1}{4k^2 |\eta_1|^2}, \\
Re(\cala_k \calb_k^{\ast}) &=& \frac{1}{4}\biggl[ \frac{1}{k^2 |\eta_1|^2}
\cos (2k|\eta_1|) + \frac{2}{k|\eta_1|}\sin (2k|\eta_1) \biggr],
\eea
 
the power spectrum of the gravitational wave in the late time is,
 from (\ref{eq:power}),
%as like in (\ref{eq:transpect}) 
\bea
\calp_h (k) \approx \biggl(\frac{H}{2\pi} \biggr) 
\biggl( 1 - \frac{H}{2M} \sin \biggl(\frac{2M}{H} \biggr) \biggr),
\label{eq:hlspect}
\eea
where, similarly as in (\ref{eq:eta1}), we have used  
\bea
|\eta_1| = \frac{M}{H k}.
\label{eq:hleta1}
\eea
 %The resulting spectrum  shows the same correction term
%as in (\ref{eq:transpect}). This is understood 
% because $\eta_1$ is determined in the  same way.   

Although we use the relation (\ref{eq:hleta1}) in order 
to get (\ref{eq:hlspect}),
since we do not have any information about the transition mechanism,
we will, for the time being, leave $\eta_1$
as a free parameter which is assumed to keep the information about the
transition.

Then, for  $k |\eta_1| \gg  \frac{M}{H}$, using
\bea
|\calb_k|^2 &\approx& \frac{H^2}{4M^2} k^2 |\eta_1|^2, \\
Re(\cala_k \calb_k^{\ast}) &\approx& -\frac{H^2}{4M^2}k^2 |\eta_1|^2
\cos (2k |\eta_1|),
\eea
we obtain the power spectrum at the late time
\bea
\calp_h(k) \approx \biggl(\frac{H}{2\pi}\biggr)^2 
\frac{H^2}{4M^2} k^2 |\eta_1|^2 ( 2+ \cos (2k|\eta_1|)).
\label{eq:hlspec1}
\eea
If we assume $\eta_1 \propto k^{-n}$, the resulting power spectrum 
behaves $\calp_n \propto k^{2-2n}$. In order to fit to the observations,
$n \approx 1$.

For $k |\eta_1| \ll \frac{M}{H}$,
\bea
|\calb_k|^2 &\approx& \frac{M^2}{4H^2} \frac{1}{k^2 |\eta_1|^2}, \\
Re(\cala_k \calb_k^{\ast}) &\approx& \frac{M^2}{4H^2} \frac{1}{k^2 |\eta_1|^2}
 \cos (2k |\eta_1|),
\eea
then the power spectrum is
\bea
\calp_h(k) \approx \biggl(\frac{H}{2\pi}\biggr)^2 \frac{M^2}{4H^2}
k^{-2}|\eta_1|^{-2} (2- \cos (2k |\eta_1|) ).
\label{eq:hlspec2}
\eea

Similarly for $kn|\eta_1| \gg M/H$, if we assume $\eta_1 \propto k^{-n}$,
$\calp_h \propto k^{-2+2n}$ and it is required to be $n \approx 1$ in order for 
the spectrum to be  scale invariant.
Otherwise, it may not satisfy the observations. $M/H$ can  also be constrained
from the observations.

%%%%%%%%%%%%%%%%%%%%%%%%%%%%%%%%%%%%%%%%%%%%%%%%%%%%%%%%%%%%%%%%%%%%%%%%
%%%%%%================Amplification of the GWs==============
\section{Spectrum of the relic gravitational wave}
\label{sect:gwspect}
%%%%%%%%%%%%%%%%%%%%%%%%%%%%%%%%%%%%%%%%%%%%%%%%%%%%%%%%%%%%%%%%%%%%%%%%

In this section, we  compute the number of gravitons produced  at the de Sitter-
radiation dominated phase transition 
in order to estimate the intensity
of the gravitational wave due to the effect from the trans-Planckian physics
 for the detection at the gravitational wave detectors. 

With the scale factor during the radiation dominated phase ($\eta_e < \eta 
< \eta_{eq}$)
\bea
a(\eta) = \frac{1}{H\eta_e^2} (\eta - 2\eta_e),
\eea
where $\eta_e$ is the time when inflation ends, 
the mode solutions of (\ref{eq:stdeq}) are given by
\bea
u_k(\eta) = \frac{\alpha_k}{\sqrt{2k}} e^{- ik\eta} 
+ \frac{\beta_k}{\sqrt{2k}} e^{ ik\eta},
\label{eq:radsol}
\eea
and the coefficients are satisfied the normalization condition
\bea
|\alpha_k|^2 - |\beta_k|^2 = 1.
\eea

Requiring the mode solutions (\ref{eq:radsol}) during the radiation dominated
phase and (\ref{eq:infsol}) during the inflationary period to be continuous
at $\eta = \eta_e$
gives
\bea
\alpha_k &=&  \cala_k  \biggl( 1- \frac{i}{k\eta_e}
-\frac{1}{2k^2 \eta_e^2} \biggr)
 + \calb_k \frac{1}{2 k^2 \eta_e^2} e^{2ik\eta_e},  \\
\beta_k &=& \cala_k \frac{1}{2 k^2 \eta_e^2} e^{-2ik \eta_e}
+ \calb_k \biggl( 1 + \frac{i}{k\eta_e} 
-\frac{1}{k^2 \eta_e^2} \biggr).
\label{eq:coeff1}
\eea

The spectrum of the relic gravitational waves is expressed in terms 
of the number of gravitons produced at the transition \cite{Maggiore:1999vm}
\bea
\Omega_{gw}(f) = \frac{1}{\rho_c}\frac{d\rho_{gw}}{d\log f}
= \frac{16\pi^2 f^4}{\rho_c} N_k, 
\eea
where $\rho_{gw}$ is the energy density of the stochastic background
gravitational waves, $f$ is the physical frequency and $\rho_c =
3H_0^2 \mpl^2/(8\pi)$ is the  critical energy density today.

From (\ref{eq:coeff1}), $N_k$ is given by
\bea
N_f &=& |\beta_k|^2 = \frac{1}{4k^4 \eta_e^4} 
+ |\calb_k|^2 \biggl( 1- \frac{1}{k^2 \eta_e^2} 
+ \frac{5}{4k^4 \eta_e^4} \biggr)  \nonumber \\
& & +\frac{1}{k^2 \eta_e^2} Re \biggl[ e^{-2ik\eta_e}
\cala_k \calb_k^{\ast} \biggl(1 - \frac{i}{k\eta_e} - \frac{1}{k^2 \eta_e^2}
\biggr) \biggr],
\eea
where we have used $|\cala_k|^2 - |\calb_k|^2 = 1$.
Since $k_{phys} = \frac{k}{a_0}
= 2\pi f$, we can express $k|\eta_e|$ in terms of the physical frequency
observed today $f$
\bea
k|\eta_e| &=& \frac{f}{f_e},
\eea
where
\bea
f_e = \frac{H_e}{2\pi(1+z_{eq})}\biggl(\frac{t_e}{t_{eq}}\biggr)^{1/2}
\simeq  10^9 \biggl( \frac{H_e}{10^{-4}\mpl }\biggr)^{1/2} \text{Hz}.
\label{eq:fe}
\eea
Here we have chosen the reference values for $H_e$ as $10^{-4} \mpl$
and taken in order to reach to the numerical values in (\ref{eq:fe})
\bea
t_e &=& \frac{1}{2H_e}, \quad
1+z_{eq} \simeq 2.4 \times 10^4 \Omega_0 h_0^2, \nonumber \\
t_{eq} &\simeq& 4.1 \times 10^{10} \Omega_0^{-2} h_0^{-4} s \nonumber \\
M_p &=& 1.22 \times 10^{19} GeV = 1.85 \times 10^{43} \text{Hz}. \nonumber
\eea

Thus, the resulting spectrum of the relic gravitational wave can be
written as
\bea
h_0^2 \Omega_{gw} 
&=& 1.2 \pi^3 \times 10^{-14} \biggl(\frac{H_e}{10^{-4}\mpl}\biggr)^2
\biggl(\frac{f^4}{f_e^4} \biggr)N_k.
\eea

Similarly, $k|\eta_1|$ can be expressed by
\bea
k|\eta_1| = \frac{f}{f_1}, 
\eea
where 
\bea
f_1 &=& \frac{H_1}{2\pi (1+z_{eq})} e^{-\caln_1} \biggl( 
\frac{t_e}{t_{eq}}\biggr)^{1/2}  \\
&=& e^{-\caln_1} \frac{H_1}{H_e} f_e
\eea
and $\caln_1 = \int_{t_1}^{t_e} H dt$.
If we assume $H_1 \simeq H_e$ during inflation , since $\caln_1 \gg 1$,
we have always $ f_1 \ll f_e$.

The modes having frequencies with $f < f_1$  exit the horizon during the 
trans-Planckian regime and the modes with $f_1 < f < f_e$ cross outside
the horizon during the inflationary period in which the linear dispersion
relations hold ($ |\eta_e| < |\eta| < |\eta_1|$).
Since the characteristic frequency corresponding to the radiation-matter
equal time ($\eta_{eq}$) is $f_{eq} \simeq 10^{-16} \Omega_0 h_0^2 \text{Hz}$,
the modes that have a frequency $f > f_{eq}$ today re-enter the horizon
during the radiation dominated phase and the modes with $f < f_{eq}$
during the matter dominated phase. In this paper, we  focus on the
frequencies with $f > f_{eq}$.

Since it is only meaningful to consider $ f \ll f_e$,  we can 
approximated the produced number of particles as
\bea
\frac{f^4}{f_e^4} N_k \simeq \frac{1}{4} + \frac{5}{4}|\calb_k|^2
- Re(\cala_k \calb_k^{\ast}).
\eea
For the usual standard inflation scenario with an adiabatic initial vacuum
($\cala_k = 1,~ \calb_k = 0$), since $N_k \propto f^{-4}$, the relic 
gravitational wave shows the flat spectrum.
Even if we consider the effect of the trans-Planckian regime, if $\eta_1$
is given by (\ref{eq:eta1}), as in (\ref{eq:transpect}) the relic spectrum
of the gravitational wave is independent of $f$.

Taking into account of the ignorance of the transition process around 
at $\eta = \eta_1$ as in HL gravity (Sect. \ref{sect:gwhl}),
 in this section we will leave $f_1$ as a free parameter.

If we define
\bea
f_c \equiv \biggl(\frac{2}{\sqrt{b_m}}\biggr)^{1/m} \frac{M}{H} f_1,
\label{eq:fc}
\eea
for $ f \gg f_1$, using the asymptotic form of $H_{\nu}^{(2)}$, 
(\ref{eq:asymphankel}), we obtain
\bea
|\calb_k|^2 &=& \frac{1}{8}\biggl(\frac{f_c}{f}\biggr)^m
\biggl[ 1 + 4 \biggl(\frac{f}{f_c}\biggr)^{2m} \biggr], \\
Re(\cala_k \calb_k^{\ast}) &=& \frac{1}{8}\biggl(\frac{f_c}{f}\biggr)^m
\cos\biggl(\frac{2f}{f_1}\biggr) 
\biggl[1-4 \biggl(\frac{f}{f_c}\biggr)^{2m}\biggr].
\eea

Hence, the spectrum for $f \gg f_c$ is 
\bea
h_0^2 \Omega_{gw}  &\simeq& 1.85 \times 10^{-13}\biggl(\frac{H_e}{10^{-4}\mpl}
\biggr)^2 \biggl(\frac{f}{f_c}\biggr)^m 
\biggl[ \frac{5}{4} + \cos \biggl(\frac{2f}{f_1}\biggr) \biggr]
\label{eq:relicspec1}
\eea
and for $f \ll f_c$
\bea
h_0^2 \Omega_{gw}  &\simeq& 4.6\times 10^{-14} \biggl(\frac{H_e}{10^{-4}\mpl}
\biggr)^2 \biggl(\frac{f_c}{f} \biggr)^m
\biggl[\frac{5}{4} - \cos \biggl(\frac{2f}{f_1} \biggr) \biggr].
\label{eq:relicspec2}
\eea
As a consequence, for $f \gg f_1$, 
the shape of the spectrum changes at $f = f_c$. If we assume 
$f_1 \propto f^n$ as in Sect. \ref{sect:gwhl}, 
the spectrum grows as $f^{m(1-n)}$ if $f > f_c$ and decreases
as $f^{-m(1-n)}$ if $f< f_c$. $m = 0$ or $n = 1$ gives a flat spectrum.
Note that the expression (\ref{eq:relicspec2}) is, in fact, not 
exact form, because, for $f_1 \ll f \ll f_c$, $z_1 \propto
(f/f_c)^m f/f_1$ and then the asymptotic form of the Hankel function
(\ref{eq:asymphankel}) is not appropriate for this range. The exact
result can be found numerically, which is shown in Fig. \ref{fig1}.

For $f \ll f_1$, the Hankel function is approximated as 
\bea
H^{(2)}_{\nu}(z) &\simeq& \frac{1}{i\sin \pi\nu}
\biggl[\frac{e^{i\nu\pi}}{\Gamma(1+\nu)}\biggl(\frac{z}{2}\biggr)^{\nu}
-\frac{1}{\Gamma(1-\nu)}\biggl(\frac{z}{2}\biggr)^{-\nu} \biggr], \quad
\text{if} ~ z \ll 1.
\eea
Since $z_1 \ll 1$, $z^{-\nu}$ term is dominant over
$z^{\nu}$ term
and then we have
\bea
|\calb_k|^2 &\propto& \biggl(\frac{M}{H}\biggr)^{\frac{3m}{2(1+m)}}
\biggl(\frac{f_1}{f}\biggr)^{\frac{9}{2}}, \\
Re(\cala_a \calb_k^{\ast}) 
&\propto& \biggl(\frac{M}{H}\biggr)^{\frac{3m}{2(1+m)}}
\biggl(\frac{f_1}{f}\biggr)^{\frac{9}{2}}.
\eea
Using these coefficients, the spectrum becomes
\bea
h_0^2 \Omega_{gw} &\propto& \biggl(\frac{H_e}{10^{-4}\mpl}
\biggr)^2 \biggl(\frac{M}{H}\biggr)^{\frac{3m}{2(1+m)}}
\biggl(\frac{f_1}{f}\biggr)^{\frac{9}{2}}.
\eea
The spectrum for $f \ll f_1$ decreases $f^{-9(1-n)/2}$ which is independent of
$m$, where we have assumed $f_1 \propto f^n$.

\begin{figure}
\centering
\includegraphics[width=0.65\textwidth]{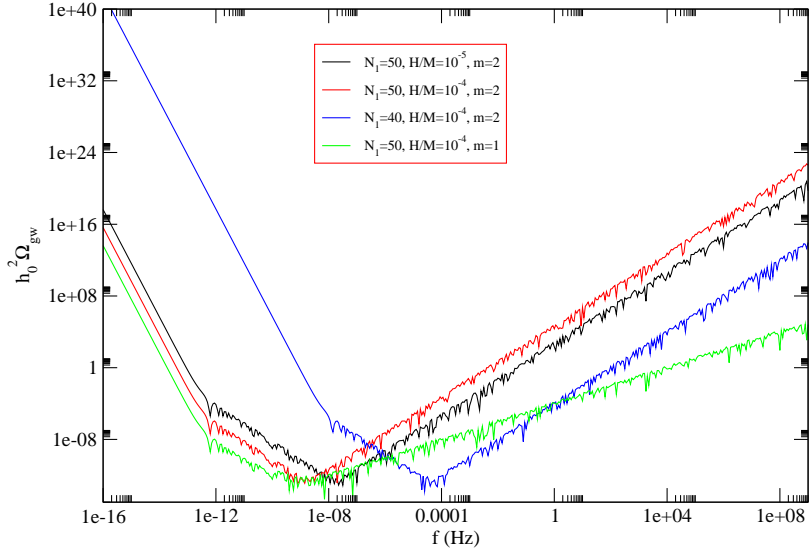}
\caption{$h_0^2\Omega_{gw}$ versus $f$ for different values of $\caln_1$, $H/M$ 
and $m$}
\label{fig1}
\end{figure}

In Fig. \ref{fig1}, we plot the spectrum of the relic gravitational wave,
$h_0^2 \Omega_{gw}$, with the physical frequency observed today
for  different values of $\caln_1$,
 $H/M$ and $m$. $m = 2$ corresponds to HL gravity. 
The graph inflects at $f = f_c$ which is given in 
(\ref{eq:fc}) and also bends at $f = f_1$.
If we take $\caln_1 = 50$ and  $H/M=10^{-5}$, $f_c \simeq 10^{-8}\text{Hz}$.
As expected from (\ref{eq:relicspec1}), 
if $f > f_c$, the spectrum increases with $f^m$ as $f$ increases. 
And if $f_1 < f< f_c$, it decreases for increasing 
$f$, but the slope is insensitive to the $m$ as explained 
in below (\ref{eq:relicspec2}).  For $f< f_1$, the spectrum decreases
with $f^{-9/2}$, independently of $m$,  as $f$ increases. 

This figure seems to indicate the violation of the cosmological bounds 
from nucleosynthesis or CMB in the low frequency range \cite{Maggiore:1999vm}
 or the present ground-base gravitational wave detector (LIGO, VIRGO etc.)
in the range $ 10\text{Hz} < f < 10^3 \text{Hz}$. Even if $m=0$,  
since $h_0^2 \Omega_{gw} \propto f^{-9/2}$ in the low frequency range, 
it might violate the cosmological bounds unless
$\eta_1$ is constrained by (\ref{eq:eta1}). This implies that
if $f_1 \propto f^{n}$ 
where $n$ should be determined from the transition mechanism,
we can expect $ n \approx 1$ in order to satisfy the experimental
and cosmological bounds.

\begin{figure}
\centering
\includegraphics[width=0.65\textwidth]{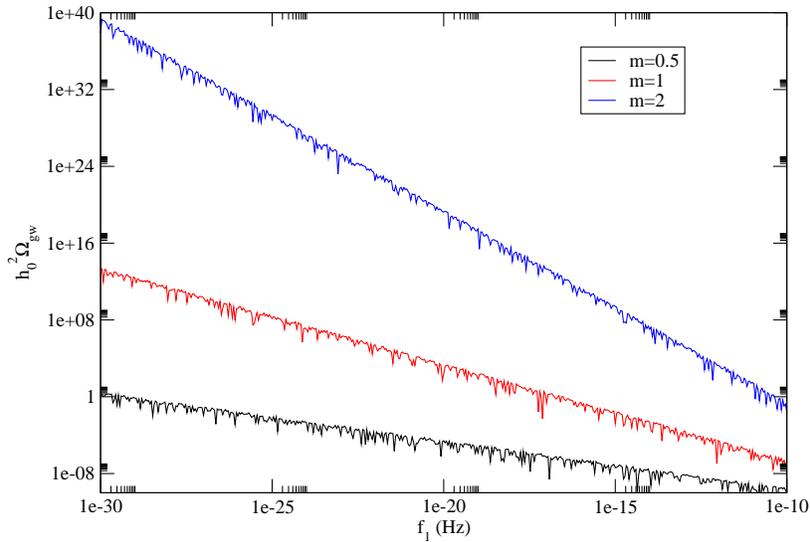}
\caption{$h_0^2\Omega_{gw}$ versus $f_1$ with $H/M=10^{-5}$ for 
$f=100 \text{Hz}$.}
\label{fig2}
\end{figure}

In Fig. \ref{fig2}, we plot $h_0^2 \Omega_{gw}$ as a function of $f_1$ at 
fixed $f = 100 \text{Hz}$ which is the frequency range of the
ground based gravitational wave detectors. The spectrum decreases
for increasing $f_1$ independently of $m$.
For a given $f_1$, as $m$ increases, the amplitude of the spectrum
also increases. As a result, $f_1$ could be constrained from the experiment,
if the gravitational waves are detected.

%%%%%%%%%%%%%%%%%%%%%%%%%%%%%%%%%%%%%%%%%%%%%%%%%%%%%%%%%%%%%%%%%%%%%%%%
%%%%%%%%%%%%%%%%%%%%%%% Summary  %%%%%%%%%%%%%%%%%%%%%%%%%%%%%%%%%%%%%%%%
\section{Summary and discussion}
\label{sect:summary}
%%%%%%%%%%%%%%%%%%%%%%%%%%%%%%%%%%%%%%%%%%%%%%%%%%%%%%%%%%%%%%%%%%%%

We have estimated the spectrum of the relic gravitational wave 
if the modified dispersion relations in the trans-Planckian regime
is given by the Corley-Jacobson dispersion relations.
The Corley-Jacobson type dispersion relations 
 could be obtained from the recently suggested quantum gravity 
model called Ho\v{r}ava-Lifshitz gravity.
Ho\v{r}ava-Lifshitz gravity  was proposed
as a model to be renormalizable quantum gravity theory in the UV region and
Einstein gravity is expected to be recovered at the IR limit. 
$m=2$ in the Corley-Jacobson dispersion relations can be considered as
HL gravity.

In obtaining the spectrum of the relic gravitational wave due to the
effect of the trans-Planckian physics, 
HL gravity phase is assumed to be followed by the inflationary 
phase in Einstein gravity and how to occur the transition
seems to play an important role. As in  many literatures dealing with
the trans-Planckian effect, if the instantaneous transition time 
is chosen by the cut-off momentum, the resulting spectrum in the late time
 is scale invariant and leads to the correction term by $H/M$. 
Since it is, however,  not clear how the transition from HL gravity
in the UV regime to the IR limit {\it i.e.} Einstein gravity takes place,
we leave $f_1$, the characteristic transition frequency,
 or $\eta_1$, the characteristic transition time, 
 as a free parameter which is assumed
to keep the information about the transition.
We have found that
in order to satisfy the experimental bounds and cosmological bounds from the
nucleosynthesis or CMB, if we assume $f_1 \propto f^{n}$ 
or $\eta_1 \propto k^{-n}$,  $n \approx 1$.

Although it is still controversial if  HL gravity is a complete theory, 
it would be interesting to investigate whether HL gravity as like any other
trans-Planckian physics can provide any detectable  effects
on the observations.  And if HL gravity
can provide the seeds for the large scale structure without inflation
as well as solve the problems of the standard Big Bang cosmology, 
it would be also interesting to find some observational signature due 
to the tensor perturbations as well as the 
scalar perturbations in HL gravity without inflation
\cite{koh09}.

\acknowledgements
The author would like to thank S. Mukohyama and G. Cristiano
 for useful comments.

%%%%%%%%%%%%%%%%%%%%%%%%%%%%%%%%%%%%%%%%%%%%%%%%%%%%%%%%%%%%%%%%%%%
%%%%%%%%%%%%%%%%%%%%%%%%%%%%%%%%%%%%%%%%%%%%%%%%%%%%%%%%%%%%%%%%%%%
%%%%%%%%%%%%%%%%%%%%%%%%%%%%%%%%%%%%%%%%%%%%%%%%%%%%%%%%%%%%%%%%%%%%%

\end{document}